\documentclass[superscriptaddress,twocolumn,floatfx,amsmath,amssymb,pra]{revtex4-2}
\usepackage{amsmath,amssymb}
\usepackage{graphics}
\usepackage{epsfig}

\usepackage{ulem}
\usepackage{MnSymbol}
\usepackage{mathrsfs}
\usepackage[usenames]{color}
\usepackage{appendix}

\begin{document}

\title{Spin hierarchy in van der Waals molecule formation via ultracold three-body recombination}

\author{Jing-Lun Li}	
\affiliation{Institut f\"{u}r Quantenmaterie and Center for Integrated Quantum Science and
	Technology IQ$^{ST}$, Universit\"{a}t Ulm, 89069 Ulm, Germany}
\author{Paul S. Julienne}
\affiliation{Joint Quantum Institute, University of Maryland, and the National
Institute of Standards and Technology (NIST), College Park, MD 20742, USA}
\author{Johannes Hecker Denschlag}
\affiliation{Institut f\"{u}r Quantenmaterie and Center for Integrated Quantum Science and
	Technology IQ$^{ST}$, Universit\"{a}t Ulm, 89069 Ulm, Germany}
 \author{Jos\'{e} P. D'Incao}
\affiliation{JILA, NIST, and the Department of Physics,
University of Colorado, Boulder, CO 80309, USA}	
\date{\today}

\begin{abstract}
 We theoretically investigate the product-state distribution of weakly bound diatomic van der Waals molecules via ultracold 
three-body recombination of bosonic alkali atoms. We find a two-level hierarchy of spin propensity rules at zero magnetic field. The primary propensity rule states that nearly all molecular products conserve the total hyperfine spin of reactant atomic pair, while molecular products not conserving the total spin are highly suppressed. For the dominant molecular products, there is a secondary propensity to conserve certain spin components of the reactant pair such as the atomic hyperfine spins, or the total electronic or nuclear spins. The second propensity varies across species and depends fundamentally on the interplay between effective electronic spin exchange and hyperfine interactions.
The spin sensitivity of product-state distribution can potentially open up 
new avenues for controlling state-to-state reaction rates in ultracold three-body recombination. 
\end{abstract}

\maketitle
\section{introduction}
Despite the intrinsically complex microscopic properties of the interatomic interactions and symmetries \cite{Collins:2002,Truhlar:2007,Qu:2018}, chemical reactions are often governed by unexpectedly general and simple fundamental principles, conservation laws and propensity rules \cite{berry1966JCP,Moore1973,Fano1985,Busto2019,Lee1991}.
The ability to study in detail the likelihood of obtaining a particular product state from well-defined chemical reactants is crucial for advancing our knowledge and uncovering novel principles and mechanisms controlling chemical reactions. 
Due to recent  advancement 
in ultracold atomic and molecular gases, experiments can now prepare reactants in a well-defined quantum 
state and detect products resolving all quantum degrees of freedom including vibration, 
rotation, electronic spin, and nuclear spin. This state-to-state resolution of chemical reactions allows for in-depth investigations of novel propensity rules in an unprecedented level of detail 
\cite{Hu2021,Wolf:2017,Wolf:2019,Haze:2022,Haze2023,Haze2024D,Hermsmeier2021,Tscherbul2006}. 
Understanding such fundamental reaction principles is crucial for developing 
control over chemical reactions and their products \cite{yang2019Sci,bigagli2023NatPhys,son2022Sci,chen2023Nat,liu2024Sci,luke2024FD}.

In ultracold atomic gases a key exothermic chemical reaction is three-body recombination, where three atoms 
collide to form a diatomic molecule and a free atom.
This chemical reaction, with rate constant $L_3$, drastically limits the lifetime and stability of Bose-Einstein condensates 
\cite{anderson1995Sci,davis1995PRL,bradley1997PRL,burt1997PRL,
inouye1998NT,courteille1998PRL,stenger1999PRL,roberts2000PRL,marte2002PRL,weber2003Sci,weber2003PRL} 
and has been extensively used as a probe to explore fundamental few-body phenomena, such as the Efimov effect \cite{braaten:2006,Naidon2017,Greene:2017,Dincao:2018}.
In recent years, detailed insights on recombination have been gained by examining its molecular product-state distribution in ultracold Rb gases 
\cite{Wolf:2017,Wolf:2019,Haze:2022,Haze2023}. 
 Such studies have found that the most populated molecular products are weakly bound molecules, 
conventionally referred to as van der Waals (vdW) molecules \cite{Howard1993, Gao:1998,Gao:2000,Gao:2004}.
 Over decades, the formation and reactions involving vdW molecules have been extensively studied across a wide range of physical chemistry processes \cite{Blaney1976ARPC,Jeziorski:1994,Wormer:2000,Koperski:2002,Reilly:2015,Hermann:2017,Brahms:2010,Brahms:2011,Quiros:2017,Mirahmadi:2021,Mirahmadi:2021c}.
Recent studies~\cite{Wolf:2017,Wolf:2019,Haze:2022,Haze2023} have revealed that, at zero magnetic field, the 
Rb$_2$ vdW molecules produced via recombination possess the same hyperfine spins as the initial atoms and that the corresponding state-to-state reaction rates are roughly proportional to the inverse of their binding energy, $1/E_b$. 
{Nevertheless, a deeper understanding of such propensity rules is 
necessary to assess their validity among other atomic species with vastly 
different physical properties. In particular,
the role of the intricate relationship between the long- and short-range physics in the spin dynamics of three-body recombination remains unclear.}
At large 
distances, hyperfine and vdW interactions govern the spin dynamics while at short
distances the dynamics is dominated by the electronic spin exchange interactions.
The interplay between hyperfine and electronic spin exchange interactions, for instance, determines the spin mixing in vdW molecules, potentially influencing how such molecular states are formed via recombination.

In this work we present a theoretical study on the product-state distribution of 
vdW molecules formed by ultracold three-body recombination to characterize the role of the molecular spin in the reaction dynamics. We consider the bosonic alkali atoms ($^{7}$Li, $^{23}$Na, $^{39}$K, $^{41}$K, 
$^{85}$Rb and $^{87}$Rb and $^{133}$Cs) at zero magnetic field. 
We find a two-level hierarchy of spin propensity rules. The first rule states that nearly all molecular products conserve
the total hyperfine spin of the reactant atomic pair.
The rates for such reactions
are generally consistent with the $1/E_b$ propensity rule \cite{Haze2023}.
In contrast, the formation of molecules not conserving the total hyperfine spin is highly suppressed and violates the $1/E_b$ propensity rule. As we will show, this points to a different formation process than that of the dominant molecular states. 
The second propensity rule expresses the likelihood
to conserve certain spin components of the 
reactant pair such as the atomic hyperfine spins, or the total electronic or nuclear spins. Although the full microscopic understanding of the origin of such spin propensity rules is elusive due to the highly complex and non-pertubative nature of the multichannel three-body interactions, we are able to introduce a dimensionless parameter $\xi_{\rm ex}$ which indicates 
what propensity to expect for a given atomic species. Such understanding is supported by  three-body numerical calculations.
Here, $\xi_\mathrm{ex}$ is the ratio of the effective electronic spin exchange and effective hyperfine interactions of the atomic pair and is an 
improved generalization of the parameter $\rho$ previously introduced in \cite{Haze:2022}.
It determines the spin structure of near-threshold molecular states and varies strongly for different atomic species. This parameter will be briefly introduced in later discussions and for more details we refer to Ref. \cite{Li2024D}. 
While for weak electronic spin exchange species, $\xi_{\rm ex}\ll1$, our findings are consistent with our previous studies of Rb atoms \cite{Wolf:2017,Wolf:2019,Haze:2022,Haze2023}, for other atomics species with intermediate and strong electronic spin exchange interactions,  $\xi_{\rm ex}\sim 1$ and $\xi_{\rm ex}\gg1$, respectively, the conclusions (in particular the second level spin propensity) are rather different.
The rest of the paper is organized as follows: in Section \ref{theo} we give a short overview of our theoretical framework. In Section \ref{dis} we analyze our numerical results and discuss the two-level hierarchy of spin propensity rules. We summarize our main conclusion in Section \ref{con}.

\section{Theoretical framework} \label{theo}

\begin{figure}[htbp]
	\includegraphics[width=\columnwidth]{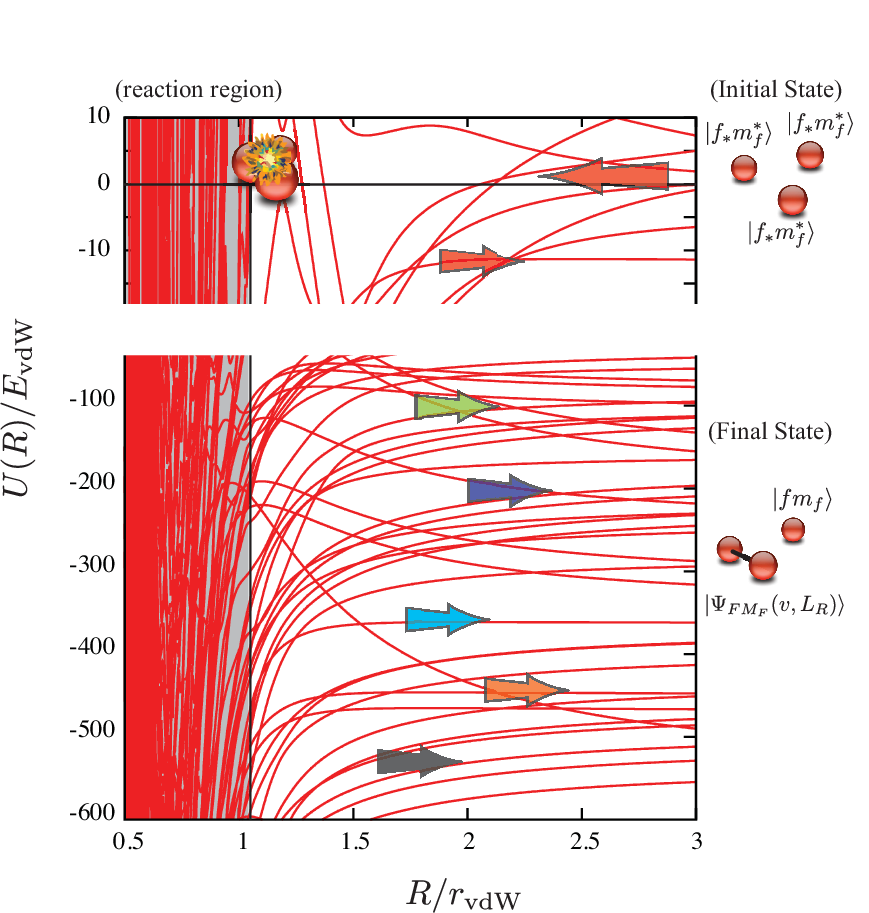}
	\caption{Three-body adiabatic potentials, $U(R)$, for $^{87}$Rb atoms in van der Waals units. 
 In three-body recombination, three free atoms, characterized by the spin product state $|f_*$=1 $m_f^*$=-1$\rangle$$|f_*m^*_f\rangle|f_*m^*_f\rangle$,
approach each other via the three-body continuum channels [i.e., channels where $U(R)$$>$$0$ for $R$$\gg$$r_{\rm vdW}$]. 
As the atoms reach the ``reaction region'' ($R\lesssim r_{\rm vdW}$ shaded area) the complex structure of
avoid-crossings drive inelastic transitions to the final atom-molecule channels [$U(R)$$<$0 for $R$$\gg$$r_{\rm vdW}$] 
characterized by the $|\Psi_{FM_F}(v,L_R)\rangle|fm_{f}\rangle$ state. The different colors of outcoming arrows indicate the variation of the spin state of molecular products.}\label{URL3}
	\label{L3Potentials}
\end{figure}

Our three-body studies are performed in the adiabatic hyperspherical representation \cite{Suno:2002,wang2011pra}, 
where the hyperradius $R$ gives the overall size of the system while the set of hyperangles $\Omega$ describe its
internal motion.
The adiabatic separation between hyperradial and hyperangular motions leads to the three-body hyperradial
Schr\"odinger equation
\begin{align}
&\left[-\frac{\hbar^2}{2\mu}\frac{d^2}{dR^2}+U_{\nu}(R)\right]F_{\nu}(R)\nonumber\\
&~~~~~~+\sum_{\nu'}W_{\nu\nu'}(R)F_{\nu'}(R)=EF_{\nu}(R),\label{Schro}
\end{align}
governing the hyperradial motion via the three-body potentials $U_{\nu}$ with inelastic transitions driven by the
nonadiabatic couplings $W_{\nu\nu'}$. The hyperangular equation is given in Appendix \ref{hyper}. Here, $E$ is the total energy, $\mu=m/\sqrt{3}$ the three-body reduced mass, 
$m$ the atomic mass, and $\nu$ the set of quantum numbers necessary to characterize each channel.
In the present study we include the atomic hyperfine structure and interatomic interactions given by the electronic singlet and 
triplet Born-Oppenheimer potentials \cite{Julienne2014,Knoop2011,Tiemann2020,Strauss2010,Berninger2013}.
Such potentials are modified to restrict their number of molecular states while still preserving the proper singlet 
and triplet scattering lengths, $a_s$ and $a_t$, respectively \cite{Chapurin:2019,Xie:2020,Li2024D}.
In Fig. \ref{URL3} we display the three-body potentials for $^{87}$Rb atoms and describe the corresponding 
process leading to three-body recombination.
We note that the vdW interaction between atoms defines the typical scale for length 
$r_{\rm vdW}=\frac{1}{2}(mC_6\hbar^2)^{1/4}$ and energy $E_{\rm vdW}=\hbar^2/mr_{\rm vdW}^2$ of the system, where $C_6$ is the vdW dispersion coefficient. Compared to this scale, our considered collisional energy of free atoms $E$ is extremely small $E\ll E_{\rm vdW}$, while the vdW molecular binding energies can be up to thousands of $E_{\rm vdW}$. In what follows we will use vdW units to allow a direct comparison between different atomic species.

In the hyperspherical representation the reaction rates are obtained from the solutions of Eq.~(\ref{Schro}), from 
which we
determine the scattering $S$-matrix \cite{wang2011pra}. The total recombination rate, $L_3$, and the corresponding 
state-to-state rates, $L_{3\beta}$, are given by \cite{Dincao:2018}
\begin{align}
    L_3=\frac{1}{2}\sum_{\alpha\beta}\frac{192\pi^2\hbar}{\mu k^4}|S_{\beta\alpha}|^2=\sum_{\beta}L_{3\beta},
\end{align}
where $k^2=2\mu E/\hbar^2$ and with $\alpha$ and $\beta$ running over 
initial (three-body continuum) and final (atom-molecule) channels, 
respectively. 
In this study, we analyze the state-to-state recombination rates 
$L_{3\beta}$ according to the binding energy and spins of the corresponding molecular product states. For all atomic species we choose the initial atomic
state to be the "spin-stretched" state of the lowest hyperfine manifold,
$|f m_f\rangle$$\equiv$$|f_* m^*_f$=-$f_*\rangle$, where $f$ is the atomic hyperfine spin and $m_f$ its azimuthal projection. Thus, we have $f_*=3$ for $^{133}$Cs, $f_*=2$ for $^{85}$Rb and 
$f_*=1$ for all others ($^{7}$Li, $^{23}$Na, $^{39}$K, $^{41}$K, and $^{87}$Rb). 
With this choice 
fast two-body losses are prevented, providing a convenient experimental condition for the study of 
product-state distribution. We note that our results are also 
valid for the state $|f_* m^*_f$=$+f^*\rangle$.

\begin{figure*}[htbp]
    \includegraphics[width=1.9\columnwidth]{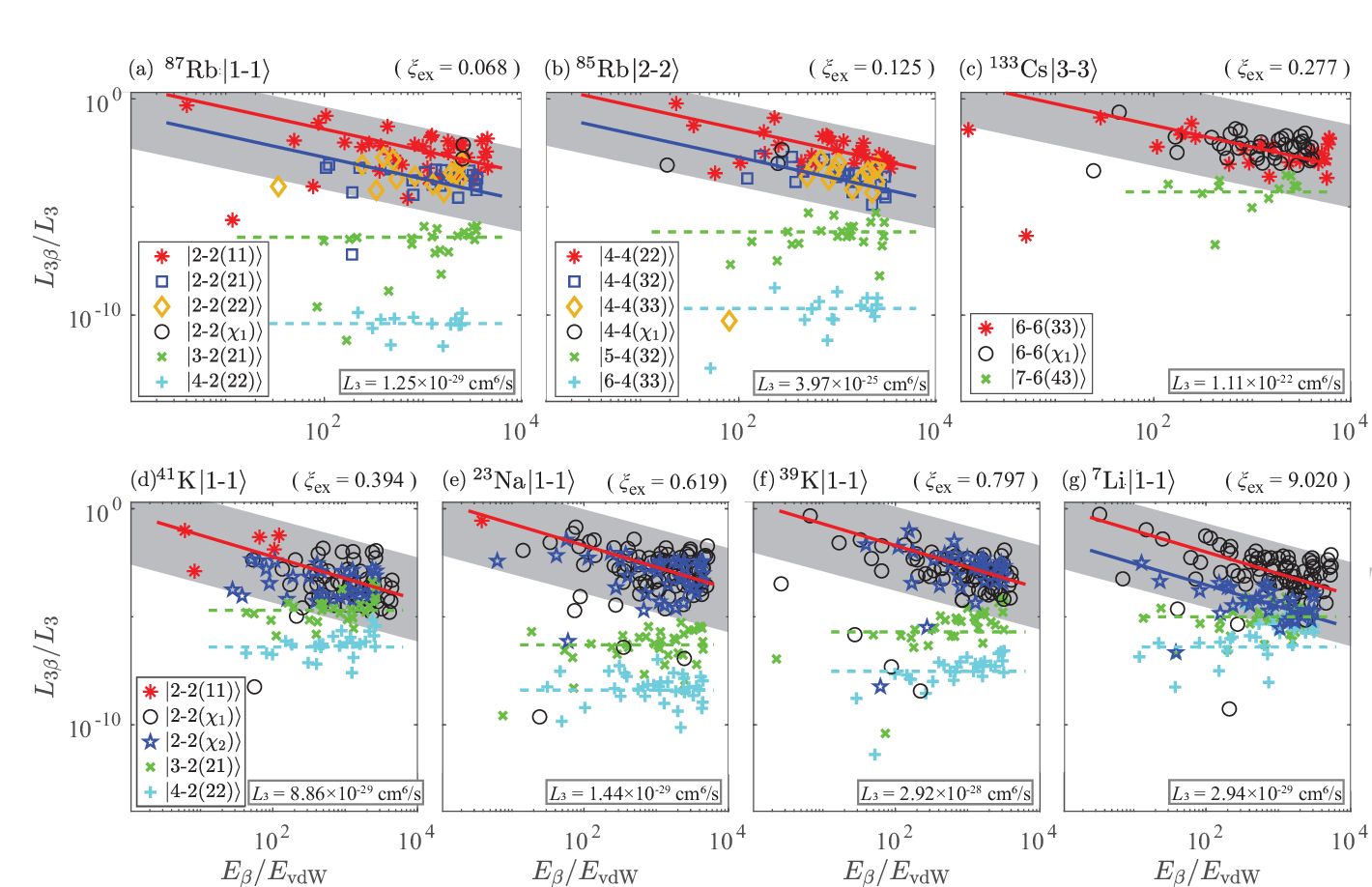}
	\caption{\label{fig:K3} Recombination fraction $L_{3\beta}/L_3$ characterizing the product-state distribution in terms of the molecular binding energies, $E_{\beta}$, and spins at vanishing magnetic field for different atomic species. 
 Here, we indicate the initial atomic spin state for each species, $|f_*m_f^*\rangle$, while the final molecular states (see legends) are 
 classified according to their molecular spin $|FM_F(f_af_b)\rangle$ 
 or, in some cases, $|FM_F(\chi_1)\rangle$ or $|FM_F(\chi_2)\rangle$ for a mixed molecular state (see Appendix \ref{spinassign} for a detail description of our assignment).
 The lower panels share the same legend. The straight solid lines 
denote the $1/E_{\beta}$ scaling of $L_{3\beta}$, while the dotted lines indicate a constant rate. The shaded area indicates the region where the partial rates typically scatter around the $1/E_{\beta}$ scaling.
For each atomic species, we indicate the corresponding values of $L_3$ and $\xi_{\rm ex}$ (see discussion in the text).}
\end{figure*}

\section{numerical analysis and discussion} \label{dis}

While the initial spin state for recombination is always a product state 
$|f_*m^*_f\rangle|f_*m^*_f\rangle|f_*m^*_f\rangle$,
the final atom-molecule state is characterized by $|\Psi_{FM_F}(v,L_R)\rangle|fm_{f}\rangle$ 
where $\Psi_{FM_F}(v,L_R)$ is the normalized total (spatial and spin) molecular wavefunction
with ($v,L_R$) representing the vibrational and rotational molecular quantum numbers \cite{Comment}. 
Our studies are performed at zero $B$-field such that the two-atom total hyperfine spin, $F$ ($|f_a-f_b|\le F\le f_a+f_b$), and its projection, $M_F=m_{f_a}+m_{f_b}$, are 
good quantum numbers for characterizing the molecular states. 
However, since in a three-body system the two-body total angular momentum quantum numbers 
$F$ and $M_F$ are not in general conserved quantities, the question is what rules control the product distribution in a dynamical process like three-body recombination.
Similar to Rb \cite{Haze:2022,Haze2023}, 
our findings across different atomic species provide clear evidence of a spin propensity 
rule favoring recombination into molecular states, $|\Psi_{FM_F}(v,L_R)\rangle$, 
whose values for $F$ and $M_F$ are the same as
that of the atomic pairs forming the initial state, thus
implying the quasi-conservation of $(F, M_F)$.
This is the first, and strongest, rule of our propensity hierarchy. 
In our case, since the initial atomic state is $|f_*m_f^*\rangle$, 
any pair of atoms will have $F_*=2f_*$ and $M_F^*=-2f_*$ and, as a result, the favorable molecular 
states are those conserving $(F_,M_F)$, i.e., with $(F,M_F)= (F_*,M_F^*)$. 

Our numerical results are shown in Fig.~\ref{fig:K3} for the recombination fraction, 
$L_{3\beta}/L_3$, for each molecular state according to their corresponding binding energies, $E_\beta$. We classify the 
states according to their molecular 
spin $|FM_F(f_af_b)\rangle$ or, in some cases, $|FM_F(\chi)\rangle$ for mixed molecular states, i.e., molecular states 
whose spins are in a superposition of $|FM_F(f_af_b)\rangle$ states of the same $F$ and $M_F$ values
(see Appendix \ref{spinassign} for more details). We note that according to the molecular spin structure of a given species, the first rule is manifest in Fig.~\ref{fig:K3} in terms of the high production rates for $|F^*M^*_F(f_af_b)\rangle$ and/or $|F^*M^*_F(\chi)\rangle$ molecular states. The molecular states included in our model \cite{Li2024D} have $(v,L_R)$ values (not shown in 
Fig.~\ref{fig:K3}) ranging from $v=-1$ to 
-7 (counting down from the most weakly bound state) and $L_R=0$ up to 26, in a total of about $100$
molecular states. We note that the data in Fig.~\ref{fig:K3} corresponding to our most deeply bound molecular 
states are likely to be model-dependent since they don't fully reproduce the actual molecular states within that 
range of energies \cite{Li2024D}.
Figure \ref{fig:K3} shows that for molecular states with $(F_*,M_F^*)$ the rates are generally consistent with the $L_{3\beta}\propto 1/E_\beta$ energy scaling propensity 
rule of Ref.~\cite{Haze2023} (see $1/E_\beta$ solid lines in Fig. \ref{fig:K3}). However, our data set encounters insufficient binding energy range and large fluctuations, making it challenging to achieve a reliable fitting. (To more accurately test the
$1/E_\beta$ propensity rule it would be necessary to include significantly more deeply bound 
molecular states, as done in the spinless model of Ref.~\cite{Haze2023}. 
Nevertheless, for our spin-dependent model this would lead to a considerable increase in the
numerical demand.)
For the molecular states whose ($F,M_F$) are not conserved, however, $L_{3\beta}$ 
is generally much smaller and depends only weakly on $E_\beta$ (see horizontal dashed lines in Fig. \ref{fig:K3}). 
We also have found that, similarly to our previous studies in Refs. \cite{Wolf:2017,Haze:2022,Haze2023}, the partial 
rates have no apparent dependence on $L_R$.
It should also be noted that 
the results in Fig.~\ref{fig:K3} were carried out by fixing the spin of the third atom at $|f_* 
m^*_f\rangle$. This restricts the three-body spin bases to those where the third atom remains in the 
incoming spin state while the other two atoms are in the allowed spin state with $M_{F}=M^*_F$, see Appendix \ref{hyper} for details. However, we have carefully examined the contribution from other spin states of the third atom 
and found it is typically below one percent. This points to a largely mechanical role of the third atom in recombination (see Appendix \ref{mutiM}).
Our result is in contrast to the strong three-body spin-exchange recombination of $^7$Li at large magnetic 
fields found in Refs. \cite{Li:2022,vandekraats:2024}. We attribute this difference to the fact that at finite 
magnetic fields $F$ and even $f$ are not good quantum numbers anymore.
A systematic study on the effects of finite magnetic fields in the 
product state distribution is beyond the scope of the present study.

We now analyse the propensity
to conserve spin components of
the reacting atomic pair. This leads to the
 second rule in the propensity hierarchy, involving the $\xi_{\rm ex}$ parameter. This parameter is the ratio of effective electronic spin exchange interaction to effective hyperfine interaction on a given molecular state, which provides a simple way to characterize the spin structure of vdW molecules. We briefly introduce the definition of $\xi_{\rm ex}$ in Appendix \ref{xi} and refer to Ref. \cite{Li2024D} for more detailed discussion.
As we will see below, 
  for very small and very large 
  $\xi_{\rm ex}$ parameters the 
  production rate depends on the 
  spin overlap of the initial atom pair and the molecular product. 

Figure~\ref{fig:K3} shows that for Rb the 
$F = F^*, M_F = F^*$ molecular states can 
be generally further characterized by the
$f_a, f_b$ quantum numbers. This is a consequence
of the small $\xi_{\rm ex} \ll 1$ parameter [see Figs.~\ref{fig:K3}(a) and \ref{fig:K3}(b)], indicating 
that for these molecular states hyperfine interaction effectively dominates over
exchange interaction (see Appendix \ref{xi}). There is a propensity to 
conserve the $f_a, f_b$ quantum numbers. 
This is because the recombination primarily occurs mechanically and requires no spin flips \cite{Haze:2022}.
Indeed, both our numerical calculations [Figs.~\ref{fig:K3}(a) and \ref{fig:K3}(b)] and experimental observations in Refs.~\cite{Wolf:2019,Haze:2022,Haze2023} have already shown this. 
If one or both atomic hyperfine spins flip
the rate  typically drops by 
 a factor $\approx 20$, indicate the difficulty of flipping atomic hyperfine spins. 

For cases of intermediate $\xi_{\rm ex}\sim1$ [as those of $^{133}$Cs, $^{41}$K, $^{23}$Na and 
$^{39}$K in Figs.~\ref{fig:K3}(c)-\ref{fig:K3}(f)] 
electronic spin exchange and hyperfine interactions are comparable and 
the $F = F^*, M_F = F^*$ molecular states are typically not well-described by the $f_a, f_b$ quantum numbers.
Their spin states are mixed, $|FM_F(\chi_1)\rangle$ or $|FM_F(\chi_2)\rangle$ (see see Appendix \ref{spinassign} for a detailed description). 
 Our results for the product state distribution for cases of intermediate $\xi_{\rm ex}\sim1$ do not show a clear separation into preferred spin states, even if other spin bases are used for characterizing the molecular products. As a result the second spin rule is violated in such cases.

For systems where $\xi_{\rm ex}\gg1$, as is the case for $^7$Li, the electronic spin exchange interaction 
dominates over the hyperfine interaction.
Most molecular states are then
well described in the $|FM_F[SI]]\rangle$ basis
where $S$ and $I$ are the total electronic and 
nuclear spin quantum numbers, respectively. 
The data for lithium are shown in 
Figs.~\ref{fig:K3}(g) and Fig.~\ref{fig:7Li} using the basis sets
$|FM_F(\chi)\rangle$
and 
$|FM_F[SI]]\rangle$, respectively, see Appendix \ref{spinassign} for spin assignment.
\begin{figure}[htbp]
	\includegraphics[width=0.85\columnwidth]{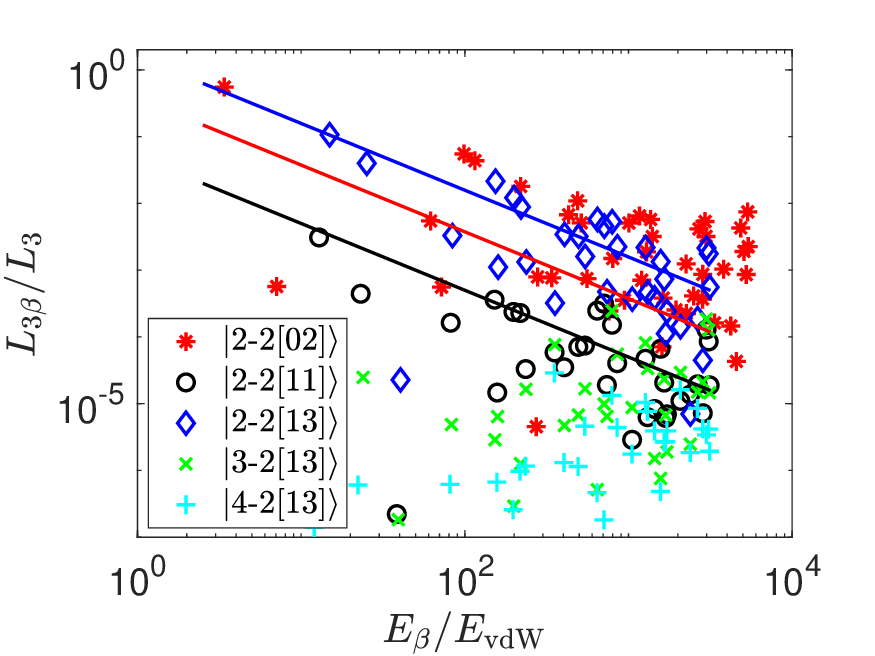}
	\caption{\label{fig:7Li} Molecular product-state distribution for three-body recombination of ultracold $^{7}$Li atoms 
 with molecular levels characterized in the $|FM_F[SI]\rangle$ basis. {The straight solid lines 
denote the $1/E_{\beta}$ scaling of $L_{3\beta}$ with prefactors in the ratio of 0.7875 : 0.1875 : 0.0250.}}
\end{figure}
We find that the molecular product rates depend upon the overlap between the initial and final 
spins, i.e., $|\langle f_*m^*_f,f_*m^*_f|FM_F[SI]\rangle|^2$.
The initial atomic scattering state $|1\text-1\rangle|1\text-1\rangle$ 
has a non-zero overlap only with the molecular states $|2\text-2[02]\rangle$, $|2\text-2[11]\rangle$, and 
$|2\text-2[13]\rangle$, with $|\langle f_*m^*_f,f_*m^*_f|F_*M_F^*[SI]\rangle|^2=$ 0.1875, 0.0250 and 
0.7875, respectively \cite{Li2024D}. 
According to this analysis the dominant molecular states should be 
$|2\text-2[13]\rangle$ and $|2\text-2[02]\rangle$, with $|2\text-2[11]\rangle$ and other spin states be only weakly 
populated. This is, in fact, consistent with our numerical calculations in Fig.~\ref{fig:7Li}, indicating a propensity to conserve the $S$ and $I$ quantum numbers.

To further test our interpretation of the spin dependence of the product-state distribution in terms of $\xi_{\rm ex}$, we have 
performed additional calculations in which we artificially change the hyperfine splitting for $^{87}$Rb atoms so that the 
corresponding $\xi_{\rm ex}$ reaches a value comparable to that of $^{23}$Na and $^{7}$Li. We find that the product-state 
distribution of three-body recombination of $^{87}$Rb changes drastically and, in fact, recovers the hierarchical 
structure of other species with similar values of $\xi_{\rm ex}$, see Appendix \ref{xi}.

As previously mentioned, the rate of formation for non-conserving $(F,M_F) \neq 
(F_*,M_F^*)$ molecular states, as shown in Fig.~\ref{fig:K3}, exhibits a weak 
dependence on $E_\beta$. This behavior contrasts sharply with that of the conserving 
$(F,M_F)$ molecular states discussed so far. In fact, at low $B$ fields, the formation of 
non-conserving ($F,M_F$) molecular states in our model can only occur via a process in which 
the spins of the atoms in the molecule are changed via the spin-exchange 
interaction with the third atom.
This is so because the spin-exchange among 
the atoms within the molecule cannot change ($F_*, M_F^*$). 
A possible process to form non-conserving ($F,M_F$) 
molecular states can be through the formation of a strongly coupled intermediate that conserves ($F,M_F$) but flips
the atomic spins ($f_a$ and/or $f_b$). This molecular state is energetically close to the final non-conserving $(F,M_F)$ molecular state, formed subsequently by interacting with the third atom. 
For $^{87}$Rb atoms [Fig.~\ref{fig:K3}(a)], for instance, the 
formation of non-conserving $|3\text-2(21)\rangle$ molecules could be accomplished via the 
intermediate state $|2\text-2(21)\rangle$ while the formation of 
$|4\text-2(22)\rangle$ molecules would be more favorable via the 
$|2\text-2(22)\rangle$ state. (Similar processes can also 
be hypothesized for other atomic species in Fig.~\ref{fig:K3}.) 
A more detailed analysis of the formation of non-conserving 
($F,M_F$) molecular states is beyond the scope of the present work. Regardless of the 
details of how such processes occur they should become more prevalent as electronic 
exchange interactions become more important. In fact, from Fig.~\ref{fig:K3} it is 
clear that the increase of the formation rates of such molecular states is consistent 
with the increase of $\xi_{\rm ex}$ across the 
different atomic species.
\section{conclusion} \label{con}

In summary, our in-depth analysis of the spin product-state distribution of vdW molecules reveals a two-level hierarchy of spin propensity rules in ultracold three-body recombination. While a complete microscopic understanding of these propensity rules remains elusive, our analysis demonstrates that they can be roughly explained using simple physical principles.
The spin hierarchy, characterizing the propensity of changes in various spins of reactant atomic pairs during molecule formation, strongly depends on the interplay between effective electronic spin exchange and hyperfine interactions across alkali species.
The demonstrated spin sensitivity of three-body recombination can potentially allow for the coherent control of this reaction via spin mixing \cite{Haze2024D,Dorer2025}. 
Such ability can, for instance, allow for the enhancement or suppression 
the population of a particular family of molecular products based on the spin 
propensity rules we uncovered in our study. 

\acknowledgments

 This work was supported by the Baden-W\"urttemberg Stiftung through the Internationale Spitzenforschung program 
(BWST, contract No.~ISF2017-061) and by the German Research Foundation (DFG, Deutsche Forschungsgemeinschaft, 
contract No. 399903135). We acknowledge support by the state of Baden-Württemberg through bwHPC
and the German Research Foundation (DFG) through grant no INST 40/575-1 FUGG (JUSTUS 2 cluster). J.H.D and J.P.D. acknowledge funding by Q-DYNAMO (EU HORIZON-MSCA-2022- SE-01) within project No. 101131418.
J.P.D. also acknowledges partial support from the U.S. National Science Foundation (PHY-2012125 and PHY-2308791) 
and NASA/JPL (1502690).

\newpage

\appendix

\section{Three-body hyperspherical representation} \label{hyper}

In the hyperspherical adiabatic representation \cite{Suno:2002,wang2011pra}
one of the major tasks is to solve the hyperangular adiabatic equation 
$\hat{H}_{\rm ad}\Phi_{\nu}(R;\Omega)=U_{\nu}(R)\Phi_{\nu}(R;\Omega)$, at fixed values of $R$, in order to determine the 
three-body potentials $U_\nu$ and channel functions $\Phi_\nu$, 
both of which are required for the study of the solutions of Eq.~(\ref{Schro}).
The adiabatic Hamiltonian $\hat{H}_{\rm ad}$,
\begin{eqnarray}
\hat{H}_{\rm ad}=\frac{\hat\Lambda^2(\Omega)+15/4}{2\mu R^2}\hbar^2+\hat{V}_T(R,\Omega)+\hat{H}_{\rm hf}(B),
\end{eqnarray}
contains the hyperangular kinetic energy via the hyperangular momentum operator \cite{Suno:2002,wang2011pra}, 
$\hat\Lambda$, the $B$-field dependent hyperfine atomic interactions, $\hat{H}_{\rm hf}$, as well as all the interatomic interactions of the system, $\hat{V}_T$. Note that at $B=0$ only the hyperfine interaction remains in $\hat{H}_{\rm hf}$.
In the present study we assume that the interaction is given by 
\begin{align}
\hat{V}_{T}(R,\Omega)=\hat{V}(r_{ab})+\hat{V}(r_{bc})+\hat{V}(r_{ca}),
\end{align}
where $r_{ij}$ is the distance between atoms $i$ and $j$. The interatomic interaction, $\hat{V}$, is represented by
\begin{align}
\hat{V}(r_{ij})=\sum_{SM_S}|SM_S\rangle V_S(r_{ij})\langle SM_S|,
\end{align}
where $S$ is the total electronic spin and $M_S$ is azimuthal projection. In our present study, we used realistic Bohr-Oppenheimer 
potentials for the singlet ($S=0$) and triplet ($S=1$) interactions for large distances but for short distances we add an additional term to control the number of bound states in 
the problem \cite{Li2024D}.

The solutions of the hyperangular adiabatic equation are obtained by expanding the channel functions $\Phi_{\nu}$ on the basis of 
separated atoms hyperfine spins states, $|\sigma\rangle\equiv|f_{a}m_{f_a}\rangle|f_{b}m_{f_b}\rangle|f_{c}m_{f_c}\rangle$,
\begin{align}
    \Phi_\nu(R;\Omega)=\sum_{\sigma}\phi^{\sigma}_{\nu}(R;\Omega)|\sigma\rangle.
\end{align}
In practice, we restrict the spin basis $|\sigma\rangle$ to those where two of the atoms (a and b) are found 
in allowed spins states for a given $M_{F}=m_{f_a}+m_{f_b}$ while the third atom (c) remains in the 
incoming spin state.
Applying this expansion to the hyperangular adiabatic equation results in a coupled system of equations for the 
components of $\phi_\nu^{\sigma}$:
\begin{eqnarray}
\left[\frac{\hat\Lambda^2(\Omega)+15/4}{2\mu R^2}\hbar^2+E^{\sigma}_{\rm hf}(B)-U_{\nu}(R)\right]\phi^{\sigma}_{\nu}(R;\Omega)\nonumber \\
+\sum_{\sigma'}V_T^{\sigma\sigma'}(R,\Omega)\phi^{\sigma'}_{\nu}(R;\Omega)=0,\label{FullAngEq}
\end{eqnarray}
where $E_{\rm hf}^{\sigma}$ is the sum of the hyperfine energies of the three separated atoms in the $|\sigma\rangle$ spin state at the magnetic field $B$.
\section{Assign molecular spin state} \label{spinassign}

We assign the spin state $|FM_F(f_af_b)\rangle$ of a given molecular state according to the projection $P_{(f_af_b)}=\int_0^{\infty}|\langle FM_F(f_af_b) |\Psi_{FM_F}(v,L_R)\rangle|^2r^2dr$.
If one projection $P_{(f_af_b)}>0.9$  
     we assign the spin state to $|FM_F(f_af_b)\rangle$ state, otherwise, they are assigned to a mixed state $|FM_F(\chi)\rangle$. To elaborate, the $|6\text-6(\chi_1)\rangle$ state of $^{133}$Cs$_2$ is a mixture of 
$|6\text-6(33)\rangle$, $|6\text-6(43)\rangle$, and $|6\text-6(44)\rangle$. The $|4\text-4(\chi_1)\rangle$ state of $^{85}$Rb$_2$ is a mixture of 
$|4\text-4(22)\rangle$ and $|4\text-4(32)\rangle$, or of $|4\text-4(22)\rangle$ and $|4\text-4(33)\rangle$. The $|2\text-2(\chi_1)\rangle$ state is a 
mixture of $|2\text-2(11)\rangle$ and $|2\text-2(22)\rangle$ for $^{87}$Rb$_2$, while it is a mixture of $|2\text-2(11)\rangle$, $|2\text-2(21)\rangle$, and $|2\text-2(22)\rangle$ for molecules of other species.
For $^{41}$K$_2$, $^{23}$Na$_2$, $^{39}$K$_2$, and $^7$Li$_2$, the mixed state $|2\text-2(\chi_2)\rangle$ is a mixture of $|2\text-2(21)\rangle$ and 
$|2\text-2(22)\rangle$. A mixed state has less than 0.1 component in the unspecified $|FM_F(f_af_b)\rangle$ state. Similarly, we assign the spin state $|FM_F[SI]\rangle$ of a given molecular state according to the projection $P_{[SI]}=\int_0^{\infty}|\langle FM_F[SI]] |\Psi_{FM_F}(v,L_R)\rangle|^2r^2dr$. If one projection $P_{[SI]]}>0.9$  
     we assign the spin state to $|FM_F[SI]]\rangle$ state. We note that one mixed  $|FM_F(\chi)\rangle$ state can also be a $|FM_F[SI]\rangle$ state for some species.

\section{Three-body calculations beyond $M_F=M^*_F$} \label{mutiM}
 \begin{figure}[t]
 \centering
  \resizebox{0.5\textwidth}{!}{\includegraphics{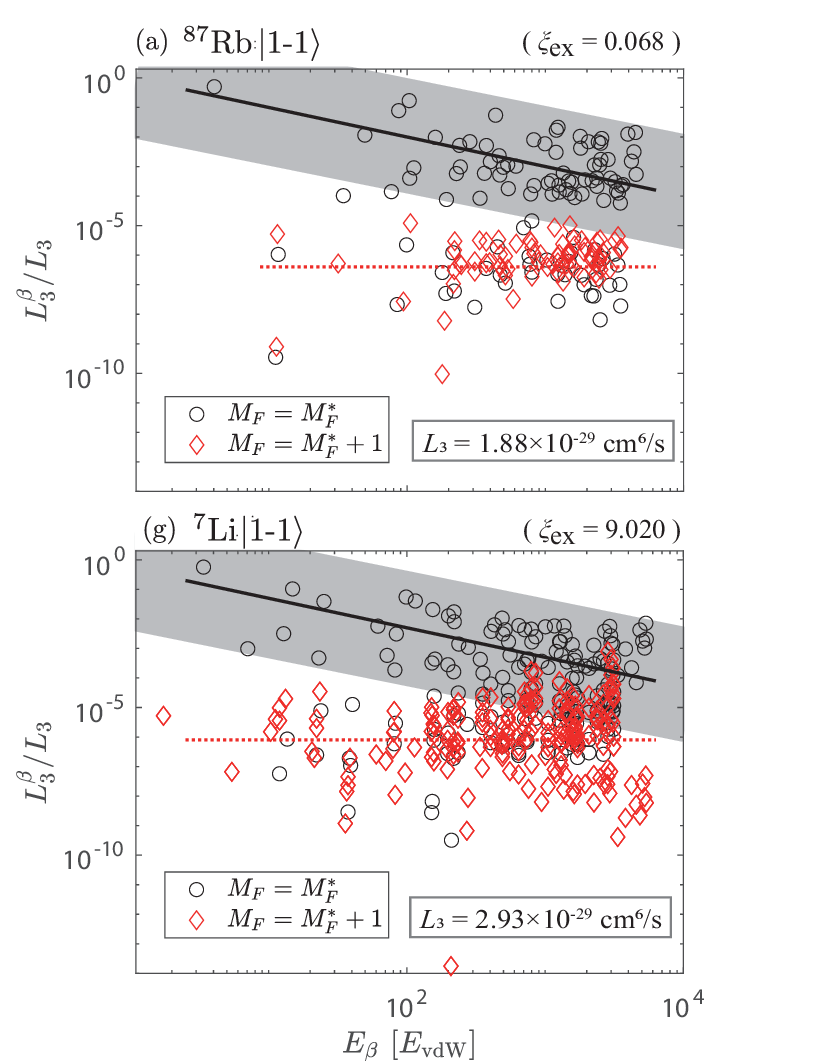} }
 \caption{\label{fig:MF} Product-state of three-body recombination for $^{87}$Rb (a) and $^7$Li (b) from numerical simulations including multiple spin states of the third atom. The straight solid lines 
denote the $1/E_{\beta}$ scaling of $L_{3\beta}$, while the dotted lines indicate a constant rate.The shadowed area indicates the typical range of scatter of the individual partial rates around the $1/E_{\beta}$ scaling lines.}
\end{figure}
Here we justify the validity of the approximation of fixing the third atom's spin state, which has been employed in most of our three-body calculations. For that, we perform additional three-body calculations by expanding the spin space of the third atom to $\{|f_* m^*_f\rangle, |(f_*+1)(m^*_f-1)\rangle\}$ so that molecular products with $M_F=M^*_F+1$ will be taken into account. Our results for $^{87}$Rb and $^7$Li using this expanded basis are shown in Fig. \ref{fig:MF}.  The calculated $L_3=1.88 \times 10^{-29}$ cm$^6$/s of $^{87}$Rb deviates from the previous value $L_3=1.25 \times 10^{-29}$ cm$^6$/s. We attribute this discrepancy to numerical fluctuations, which have negligible influence on $L_{3\beta}/L_3$. For $^{87}$Rb, the second minor discrepancy compared to the previous result [in Fig.~\ref{fig:K3}(a)] is that the recombination rates of non-conserving ($F,M_F$) molecular products are more concentrated in Fig. \ref{fig:MF}(a). This indicates again the formation of non-conserving ($F,M_F$) molecular states is sensitive to the system's interactions. Nevertheless, we find in both atomic species the main conclusions drawn from the previous single $M_F$ basis simulations do not change and the molecules with $M_F=M^*_F+1$ are weakly populated. The total reaction fraction to these molecules is below one percent for $^7$Li and on the order of 10$^{\text{-}4}$ for $^{87}$Rb. The low reaction rates into the molecular states with $M_F \neq M^*_F$ indicates that the third atom in general interacts mechanically when the other two atoms form a molecule. As discussed in three-body recombination of $^{85}$Rb \cite{Haze:2022}, the third atom should come into proximity (at a typical internuclear separation the classical out-turning point of the molecular state) with the two atoms forming a bond, such that it facilitates the release of excess energy during molecular formation. Since we consider only weekly bound vdW molecules in the present study, the spin-independent vdW interaction is more prominent than the electronic spin exchange interaction at such internuclear separations \cite{Li2024D}. As a result, when the third atom interacts, the mechanical contribution should be more dominant than the spin-flip contribution.  

\section{The exchange parameter $\xi_{\rm ex}$} \label{xi}
 \begin{figure}[t]
 \centering
  \resizebox{0.5\textwidth}{!}{\includegraphics{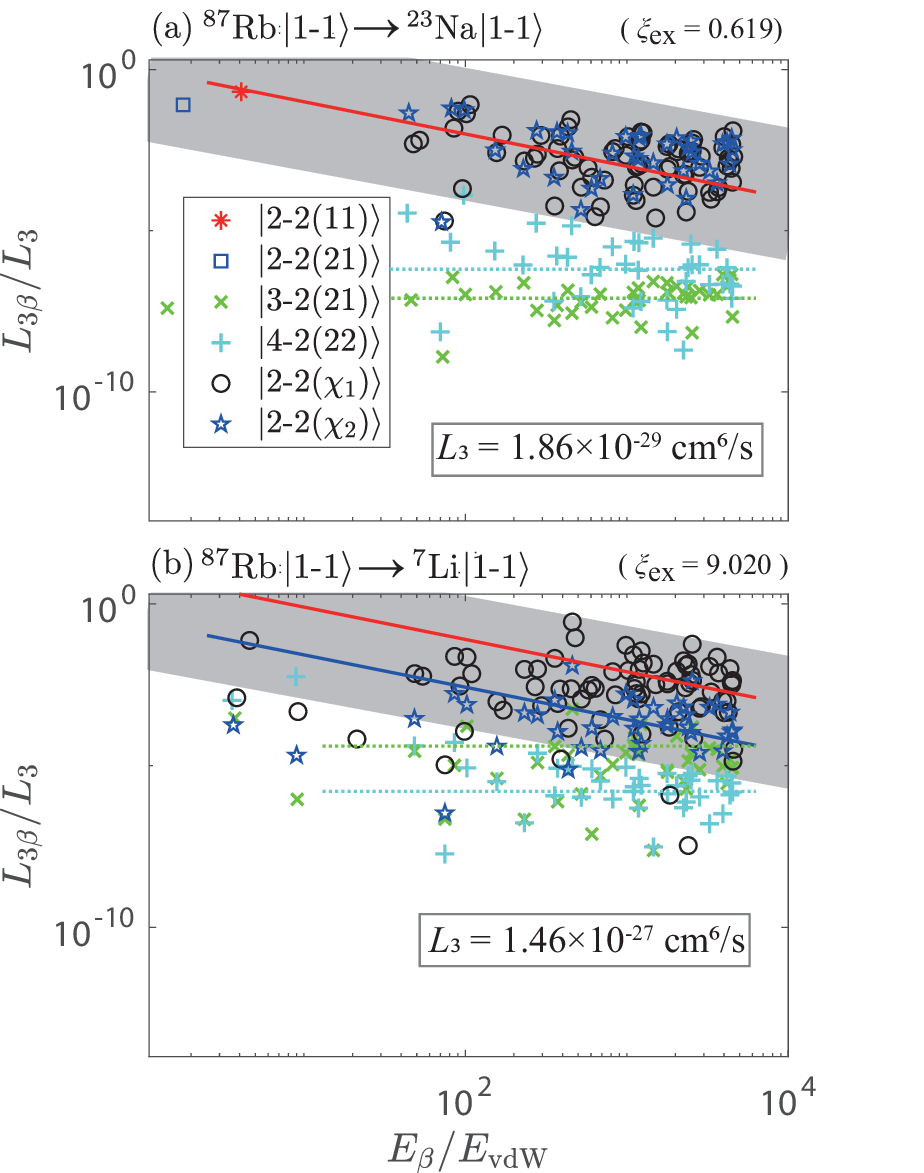} }
 \caption{\label{fig:Rb2NaLi} Product-state distribution for three-body recombination of $^{87}$Rb with artificially suppressed atomic hyperfine splitting.
 The resulting $\xi_{\rm ex}$ has the same value as that of (a) $^{23}$Na and (b) $^{7}$Li (b). The straight solid lines 
denote the $1/E_{\beta}$ scaling of $L_{3\beta}$, while the dotted lines indicate a constant rate. The shadowed area indicates the typical range of scatter of the individual partial rates around the $1/E_{\beta}$ scaling lines. (b) shares the legend with (a).}
\end{figure}

 In this section we briefly discuss the physics encapsulated in the definition of the dimensionless parameter $\xi_{\rm ex}$, presented in more details in Ref. \cite{Li2024D}. We define it as the ratio of the effective electronic spin exchange energy, $\tilde{E}_{\rm ex}$, to the effective hyperfine, $\tilde{E}_{\rm hf}$, interaction, 
\begin{equation}
\xi_{\rm ex}\equiv\frac{\tilde{E}_{\rm ex}}{\tilde{E}_{\rm hf}},
\end{equation}
 with $\tilde{E}_{\rm ex}$ and $\tilde{E}_{\rm hf}$ discussed below. This definition of $\xi_{\rm ex}$ is inspired by the parameter $\rho$ introduced in Ref. \cite{Haze:2022} and provides a simple way to characterize the spin structure of vdW molecules across different species \cite{Li2024D}.

In order to estimate the characteristic electronic spin exchange interaction we make use of the universality of the dimer spectrum for vdW interactions \cite{chin:2010, Gao:2000}. The dimer levels universally span a specific energy interval for each vibrational quantum number $v$ \cite{chin:2010, Gao:2000}. The $v=-1$ level must be in the energy range [-39.5$E_{\rm vdW}$, 0], the $v=-2$ levels in [-272.5$E_{\rm vdW}$, -39.5$E_{\rm vdW}$], and so on, regardless the short-range detail of the potentials. These universal energy intervals are known as vdW energy bins \cite{chin:2010}. The short-range detail determining the actual location of dimer levels in vdW energy bins depends on the scattering length $a$ and can be parameterized by the quantity $u$ defined as \cite{Friedrich:2004}    
\begin{equation}
u(a)=\tan^{-1}[\bar{a}/(a-\bar{a})]/\pi, \label{ua}
\end{equation}
where $\bar{a}\approx0.9560$ $r_{\rm vdW}$. For instance, $u=0$ (or $a=\pm\infty$) indicates that the levels appear at the boundaries of the energy bins. Note that as $a$ changes and the dimer energy level moves with the bin the value of $u$ varies within the range [-0.5, 0.5]. For alkali atoms, where the interaction is characterized by the singlet, $a_s$, and triplet, $a_t$, scattering lengths, the quantity $|u(a_s)-u(a_t)|$ will describe the relative position of a pair of singlet and triplet levels in each energy bin, varying between 0 and 1. 
In order consider a more general case where the nearest singlet and triplet level can belong to the same or adjacent energy bins we define $u_{st}=\min[|u(a_s)-u(a_t)|, 1-|u(a_s)-u(a_t)|]$ to characterize the relative position of such states. Here, $u_{st}$ varies from 0 to 1/2,
where whenever $u_{st}\approx0.5$, the nearest singlet and triplet energy levels are well separated indicating that the electronic spin exchange interaction between these levels is strong. For $u_{st}\ll1/2$, since energy difference between the singlet and triplet energy levels is small, the electronic spin exchange interaction is weak between these levels. According to our discussion above, 
the quantity 
\begin{equation}
\tilde{E}_{\rm ex}=u_{st}E_{\rm bin}
\end{equation}
will characterize the energy shift between the nearby singlet and triplet levels caused by the electronic spin exchange interaction. 

Similarly, we can characterize the hyperfine interaction by the energy difference between different hyperfine molecular states.
In general, the hyperfine interaction shifts the dimer levels by the atomic hyperfine splitting constant $E_{\rm hf}$ or 2$E_{\rm hf}$. Nevertheless, $E_{\rm hf}$ will not properly describe the energy difference of nearby hyperfine molecular levels when $E_{\rm hf} \gtrsim E_{\rm bin}$. Instead, this energy difference should be constrained by the size of the corresponding energy bin, typically with an upper limit around $E_{\rm bin}/2$ \cite{Li2024D}. Therefore, we use 
\begin{equation}
\tilde{E}_{\rm hf}= \min[E_{\rm hf},E_{\rm{bin}}/2]
\end{equation}
to characterize the effective hyperfine interaction. In fact, the definition above is crucial for some species, for instance $^{133}$Cs, where the hyperfine splitting is very large ($E_{\rm hf}\approx3455E_{\rm vdW}$). 

 It should be noted that for the discussion in the present work, we use the value of $\xi_{\rm ex}$ with $E_{\rm bin}=39.5$ $E_{\rm vdW}$ of $v=-1$. This choice emphasizes the nature of the most weakly bound molecular states in the system. In fact, taking the dependence of $E_{\rm bin}$ on $v$ into account for the molecular states we considered in our study, $\xi_{\rm ex}$ remains rather constant for $^{85}$Rb, $^{87}$Rb and $^{133}$Cs but increase for higher values of $v$ for $^{7}$Li, $^{23}$Na, $^{39}$K and $^{41}$K \cite{Li2024D}. As a result, our classification of $^{85}$Rb and $^{87}$Rb as atomic species with weak electronic spin exchange ($\xi_{\rm ex} \ll 1$), and $^{7}$Li as a strong electronic spin exchange species ($\xi_{\rm ex} \gg 1$) and $^{133}$Cs as intermediate electronic spin exchange species ($\xi_{\rm ex}\lesssim1$) remain unchanged. The electronic spin exchange interaction is, however, underestimated by the presented $\xi_{\rm ex}$ of $v=-1$ for $^{23}$Na, $^{39}$K and $^{41}$K. In fact, most more deeply bound (if not all) $^{23}$Na$_2$, $^{39}$K$_2$ and $^{41}$K$_2$ molecules are characterized by $\xi_{\rm ex}\gg1$ \cite{Li2024D}.

To further demonstrate the predictive power of $\xi_{\rm ex}$, we artificially change the hyperfine splitting of $^{87}$Rb while keeping the singlet and triplet scattering lengths fixed so that the new value for $\xi_{\rm ex}$ is similar to that for $^{23}$Na or $^7$Li. For the relationship between $\xi_{\rm ex}$ and the atomic hyperfine splitting and the singlet and triplet scattering lengths, we refer to Ref. \cite{Li2024D}. We find that such an artificial change in $^{87}$Rb drastically changes the product state distribution of recombination originally displayed in the main text [Fig.~\ref{fig:K3}(a)]. Despite the difference in total recombination rate $L_3$, the new $^{87}$Rb product state distribution resembles that of the real $^{23}$Na [Fig.~\ref{fig:K3}(e)] or real $^7$Li [Fig.~\ref{fig:K3}(g)], as is shown in Figs. \ref{fig:Rb2NaLi}(a) and \ref{fig:Rb2NaLi}(b), respectively. In particular, the rates for the molecular products in the $|2\text{-}2(\chi_1)\rangle$ state are more prominent than those for $|2\text{-}2(\chi_2)\rangle$ in Fig. \ref{fig:Rb2NaLi}(b) [similar to Fig.~\ref{fig:K3}(g)] while they are comparable in Fig. \ref{fig:Rb2NaLi}(a) [similar to Fig.~\ref{fig:K3}(e)]. Nevertheless, a perceivable discrepancy in the result of non-conserving ($F,M_F$) molecular states between Fig. \ref{fig:Rb2NaLi}(a) and Fig.~\ref{fig:K3}(e) indicates that the formation of these states is more sensitive to the system's interactions.


\begin{thebibliography}{68}
\bibitem [{Com()}]{Comment}%
  \BibitemOpen
  \href@noop {} {}\bibinfo {note} {Note that for simplicity our notation of the
  atom-dimer state, $|\Psi_{FM_{F}}(v,L_R)\rangle|fm_{f}\rangle$, omits the
  atom-dimer relative motion and any additional quantum numbers that
  characterize the molecular eigenstate.}\BibitemShut {Stop}%
\end{thebibliography}

%

\end{document}